\documentclass{emulateapj}

\usepackage{amsmath}
\usepackage{amssymb}
\usepackage{natbib}
\bibpunct{(}{)}{,}{}{,}{,}

\shorttitle{Excitation of incompressible waves}
\shortauthors{Morton et al.}

\begin{document}
\title{Evidence for the photospheric excitation of incompressible chromospheric waves}

\author{R. J. Morton$^{1,2}$, G. Verth$^{1}$, V. Fedun$^{1,3}$, S. Shelyag$^{4,5}$, R. Erd\'elyi$^{1}$}
\affil{$^1$Solar Physics and Space Plasma Research Centre
(SP$^2$RC), University of Sheffield, Hicks Building, Hounsfield
Road, Sheffield S3 7RH, UK\\
$^2$Mathematics and Information Sciences, Northumbria University, Newcastle Upon Tyne,
NE1 8ST, UK\\
$^3$Department of Automatic Control and Systems Engineering, University of Sheffield,
Mappin Street, Sheffield, S1 3JD, UK\\
$^4$Astrophysics Research Centre, School of 
Mathematics and Physics, Main Physics Bulding, Queen's University Belfast,
Belfast, County Antrim, BT7 1NN, UK\\
$^5$Monash Centre for Astrophysics and School of Mathematical Sciences, Monash University, Clayton, Victoria 
3800, Australia}

\email{richard.morton@northumbria.ac.uk}

\date{Received /Accepted}
\begin{abstract}
{Observing the excitation mechanisms of incompressible transverse waves is vital for determining how 
energy propagates through 
the lower solar atmosphere. We aim to show the connection between convectively driven photospheric flows and 
incompressible chromospheric waves. The observations presented here show the propagation of incompressible 
motion through the quiet lower solar atmosphere, from the photosphere to the 
chromosphere. We determine photospheric flow vectors to search for signatures of vortex motion and 
compare results to {photospheric flows present in} convective simulations. Further, we search for the 
chromospheric response to vortex motions. Evidence is presented that suggests incompressible waves can be 
excited by the vortex motions of a strong magnetic flux concentration in the photosphere. A chromospheric 
counterpart to the photospheric vortex motion is also observed, presenting itself as a quasi-periodic torsional 
motion. Fine-scale, fibril structures that emanate from the chromospheric counterpart support transverse waves 
that are driven by the observed torsional motion. A new technique for obtaining details of transverse waves from 
time-distance diagrams is presented and the properties of transverse waves  (e.g., amplitudes and periods) excited 
by the chromospheric torsional motion are measured.}
\end{abstract}


\section{Introduction}
Magnetohydrodynamic (MHD) wave phenomena have now been observed to be ubiquitous throughout the solar 
atmosphere and are considered to be a potential mechanism for the transport of energy for the heating of the solar 
atmosphere and the acceleration of the solar wind (for reviews, see, e.g., \citealp{ASC2004}; \citealp{KLI2006}; 
\citealp{ERD2006b}). However, no individual mechanism has yet been identified for {converting a
portion} of the mechanical energy generated in the Sun's convection zone to heat.

Both incompressible and compressible MHD waves are widely reported throughout the solar atmosphere (for 
reviews, see, e.g., \citealp{BANetal2007}, \citealp{ZAQERD2009}, \citealp{WAN2011}, \citealp{DEMNAK2012}) and 
incompressible waves in the solar wind (e.g., \citealp{TUMAR1995}). Recent advances in ground and space-based 
solar telescopes have allowed for {the detection of} ubiquitous incompressible transverse waves, both in the 
chromosphere (\citealp{DEPetal2007}; \citealp{HEetal2009, HEetal2009b}; \citealp{OKADEP2011}; 
 \citealp{ANTVER2011}; \citealp{KURetal2012}; \citealp{MORetal2012c}) and in the solar corona 
(\citealp{TOMetal2007}, \citealp{ERDTAR2008}). However, the incompressible waves are difficult to dissipate without 
strong gradients in Alfv\'en speed (e.g., resonant absorption - \citealp{ION1978}) 
or some process to cause a cascade of the wave energy to higher frequencies (e.g., MHD turbulence - 
\citealp{MATetal1999}) where they are efficiently dissipated by, e.g., cyclotron damping (\citealp{MCKAXF1997}).

The solar magnetic field acts as a channel with the potential to distribute wave energy around the solar atmosphere. 
In the quiet Sun at the photospheric level, a large percentage of the magnetic flux appears to be concentrated into 
intense magnetic elements (spatial scales of $100-200$~km) that outline the supergranule network. Some of these 
concentrations of magnetic flux are observed to undergo a significant expansion at the chromosphere/Transition 
Region, forming magnetic funnels that could be preferential sites for solar wind acceleration or the legs of large-
scale coronal loops (\citealp{PET2001}). The remainder of the flux may form cell-spanning chromospheric 
structures, providing a magnetic canopy (\citealp{DOWetal1986}; \citealp{RUT2006, RUT2007}; 
\citealp{WEDetal2009}). Smaller concentrations of magnetic flux also exist in the internetwork (e.g., 
\citealp{FAUetal2001}; \citealp{DOMetal2003}; \citealp{SCHTIT2003}), potentially generating additional magnetic 
structure below the magnetic canopy. 

The largest source of wave-energy is the continually evolving sub-photospheric convection, which
generates a wide spectrum of MHD fluctuations. The influence of convection
on the magnetic elements is well studied at the photospheric level. Typically, the observations have focused on
tracing the motion of strong, localised magnetic flux elements, referred to as magnetic bright points (MBPs) when 
seen in the G-band 4305~{\AA} line (e.g., \citealp{TITetal1989}, \citealp{BERetal1995}, \citealp{BERetal1998}, 
\citealp{BERTIT2001}, \citealp{SANetal2004}) or more recently in the  H$\alpha$ line wings
(\citealp{LEEetal2006}, \citealp{CHITetal2012}). The continuous jostling of the MBPs leads to the excitation of MHD 
waves and oscillations, well described by the theory of MHD oscillations in magnetic flux tubes (e.g., 
\citealp{SPR1982}; \citealp{EDWROB1983}; \citealp{ERDMOR2009}; \citealp{ERDFED2010}).


\begin{figure*}[!htp]
\centering
\includegraphics[scale=0.9, clip=true, viewport=0.0cm 0.0cm 19.5cm 17.0cm]{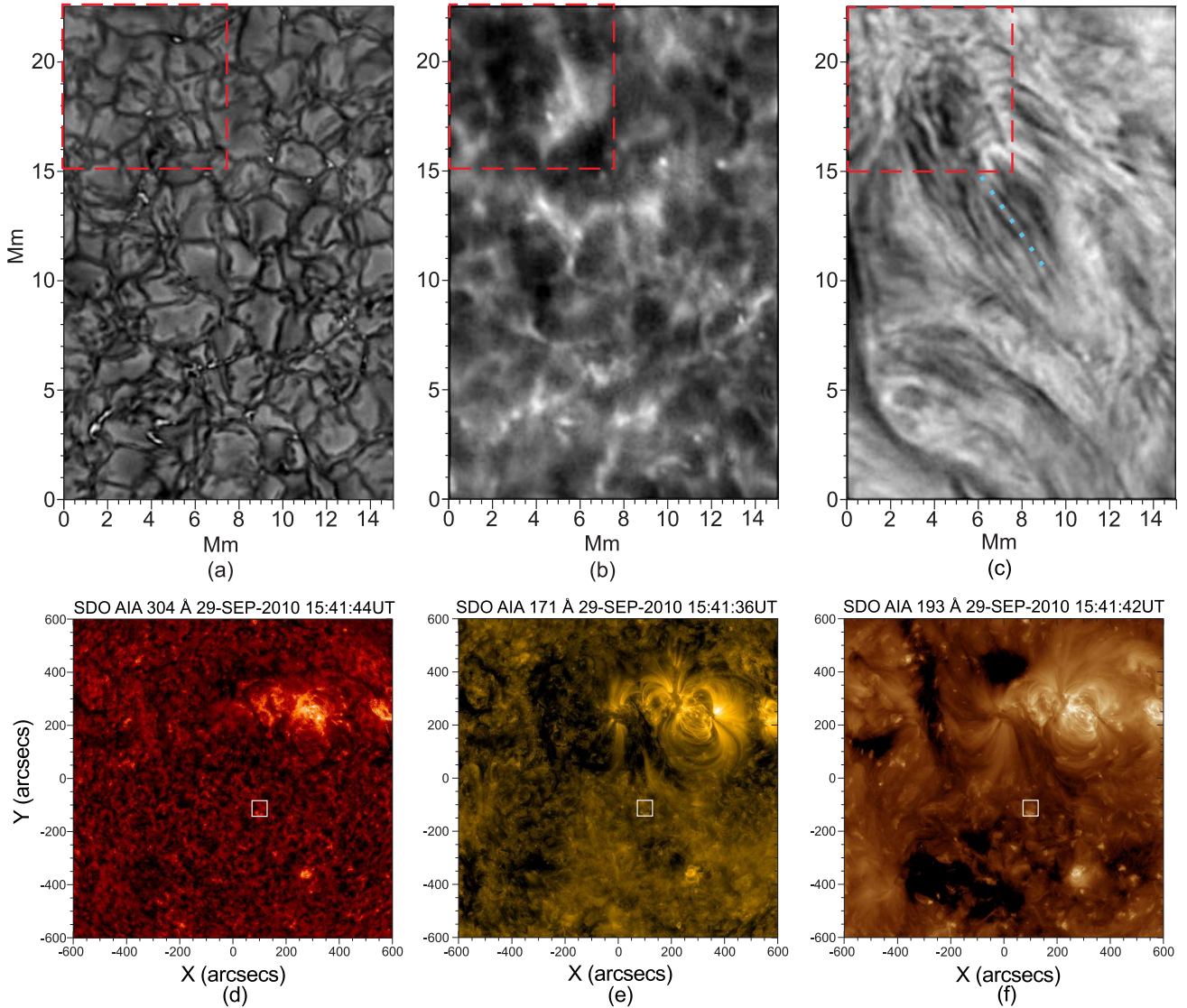} 
\caption{A region of the quiet solar atmosphere as observed by ROSA. \textit{(a)} G-band image showing a 15 by 
23~Mm$^2$ sub-region of the ROSA field of view. Collections of magnetic bright points are clearly visible in the 
inter-granule lanes. \textit{(b)} Ca K image of the lower chromosphere. \textit{(c)}
 H$\alpha$ core image of the mid to upper chromosphere. The existence of fine-scale structuring in the bandpass 
is evident, with both spicules/mottles and cell-spanning fibrils identifiable. {The blue dots show the path 
along which perpendicular cross-cuts where taken, see Section~5 for further details.} The bottom row displays 
images of the upper solar atmosphere observed by SDO/AIA. 
The panels show \textit{(d)} $304$~{\AA}, \textit{(e)} $171$~{\AA} and \textit{(f)} $193$~{\AA} filters. The axis 
show solar coordinates. The white boxes highlight the region which ROSA observed.}\label{fig:rosafov}
\end{figure*}

One particular aspect of convective motion that has received increased attention recently is the generation of 
vortices at the solar surface. Observations
(\citealp{BRAetal1988}; \citealp{BONetal2008}; \citealp{BALetal2010}) and advanced simulations 
(\citealp{MOLetal2011}; \citealp{SHEetal2011})
of solar granulation have revealed these vortex features are almost everywhere on the solar disk. 
\cite{WEDROU2009} and \cite{WEDetal2012} have
demonstrated that the vortex motion has an observable influence on the upper layers of the solar atmosphere, 
suggesting they may play a role in heating. Further, simulations of \cite{FEDetal2011b} have demonstrated that the 
twisting of an open flux-tube by vortices at the photospheric level leads to the generation of MHD waves, in 
particular incompressible (i.e., torsional Alfv\'en and transverse/kink) waves.

The transverse waves have recently been the subject of some controversy, with discussions on their nature, 
properties and nomenclature. The properties of the transverse waves in solar magnetic flux tubes are well 
understood (e.g., \citealp{EDWROB1982}; \citealp{SPR1982}; \citealp{EDWROB1983}, \citealp{ERDFED2007}), with a 
significant volume of research dedicated to developing realistic models, including, e.g., magnetic structuring 
(\citealp{VERERD2008}; \citealp{RUDetal2008}), density structuring (\citealp{ANDetal2005a}; 
\citealp{DYMRUD2005}), flux tube geometry (\citealp{VANetal2004}; \citealp{DYMRUD2006b}; 
\citealp{ERDMOR2009, MORERD2009} ), multiple flux tubes (\citealp{ROBetal2010}; \citealp{LUNetal2010} 
\citealp{PASetal2011}), resonant damping (\citealp{RUDROB2002}; \citealp{GOOetal2002}, \citealp{TERetal2006}; 
\citealp{TERetal2010c} ), partial ionization (\citealp{SOLetal2009, SOLetal2009b}; \citealp{ZAQetal2011}). 

One of the most important properties of the low-frequency transverse waves is that they are highly incompressible 
(\citealp{GOOetal2009}), requiring large gradients in the Alfv\'en speed to damp the waves. It is this property 
(shared with low-frequency bulk Alfv\'en waves of a homogeneous plasma) that means incompressible waves can 
pass through the lower solar atmosphere relatively undamped, making them a good candidate for acceleration of 
the solar wind. Observations of incompressible waves in the quiet chromosphere (\citealp{DEPetal2007}, 
\citealp{MORetal2012c}), quiet corona (\citealp{TOMetal2007}, \citealp{ERDTAR2008}) and the solar wind (\citealp{TUMAR1995}) support 
this idea. Another important property, that makes these waves relatively easy to observe, is that the transverse wave 
causes a physical displacement of the flux tubes axis.

To assess exactly what role the incompressible MHD waves play in determining the dynamics of the solar 
atmosphere, one needs to combine advance models of wave propagation throughout the atmosphere (e.g., 
\citealp{CRAVAN2005}, \citealp{CRAetal2007}, \citealp{MATSHI2010}, 
\citealp{ANTSHI2010}, \citealp{FEDetal2011}, \citealp{VANBALLetal2011}, \citealp{VIGetal2012}) with observations. 
The observations should aim to determine how the waves are being generated and track the wave-energy as it 
propagates through the solar atmosphere. This should reveal how much wave energy is present in each distinct 
region of the solar atmosphere and where the dissipation of this energy occurs. The tracking of waves through the 
solar atmosphere has been made a realistic possibility by the development of high spatial 
and temporal resolution multi-filter systems, that can observe different heights in the lower atmosphere 
simultaneously (e.g., Rapid Oscillations in the Solar Atmosphere (ROSA), CRisp Imaging Spectro-Polarimeter). The 
potential of these observational set-ups for wave tracking is demonstrated in \cite{VECetal2007}, 
\cite{JESetal2012a, JESetal2012b}. Such ground-based telescopes complement the
satellites, e.g. Hinode, Solar Dynamics Observatory (SDO), which provide detailed observations in UV/EUV. 

This paper reports the observation of both chromospheric torsional Alfv\'en and kink waves and provides evidence 
that the waves are generated by photospheric vortices. The analysed data is a high resolution, high 
cadence series of G-band, Calcium K and Hydrogen~$\alpha$ lines observed with ROSA located at the Dunn Solar 
Telescope, USA. The ROSA observations are supplemented with SDO data. Photospheric flows are determined using 
Local Correlation Tracking (LCT) and compared to numerical simulations of convection (Section~4). A statistical 
method for obtaining information on chromospheric kink waves is described based on the analysis of time-distance 
diagrams.

\section{OBSERVATIONS AND DATA REDUCTION}
The data presented here has been studied in part in \cite{KURetal2012} and \cite{MORetal2012c}. The data were 
obtained at 15:41-16:51~UT 
on 29 September 2010, with the Dunn Solar Telescope at Sacramento Peak, USA. A six-camera system called ROSA 
was employed, details of
 which are given in \cite{JESetal2010}. In brief, a $69''.3\times69''.1$ 
region of the quiet solar atmosphere, positioned close to disk centre (N0.9, W6.8), was imaged with a spatial 
sampling of 
$0''.069$~pixel$^{-1}$. We note here this is the corrected pointing to that given in \citealp{KURetal2012}. 

During the observations, high-order adaptive optics (\citealp{RIM2004}) were used to correct for wave-front 
deformations in real time. The seeing conditions were good but variable. Some frames in each of the data series are 
subject to significant distortions. We exclude periods of {seeing containing frames of inadequate quality.} 

ROSA obtained images in multiple wavelengths including G-band (4305.5~\AA - 9~{\AA} width), Ca K 
(3933.7~{\AA}) narrowband (1~{\AA}) and H$\alpha$ core (6562.8~{\AA}) narrowband (0.25~{\AA}) filters. The 
G-band data was sampled at 16.6 frame s$^{-1}$, H$\alpha$ at 2.075 frame s$^{-1}$ and Ca K 1.66 frame 
s$^{-1}$ and the images were improved by using speckle reconstruction (\citealp{WOEetal2008}) utilising a $16-1$ 
ratio. The {cadence of the reconstructed} G-band, H$\alpha$ and Ca K time-series are $0.96$~s, $7.7$~s 
and $9.6$~s, respectively. To ensure accurate co-alignment in all bandpasses, the broadband times series were 
Fourier co-registered and de-stretched  (\citealp{JESetal2007}).   

Images from the SDO Atmospheric Imaging Assembly (AIA) are also presented to 
provide context for the state of the upper atmosphere above the above the ROSA field of view. The images from 
SDO are processed with \textit{aia\_prep} and have a spatial resolution of 0''.6 pixel$^{-1}$.


\begin{figure*}[!tp]
\centering
\includegraphics[scale=0.9, clip=true, viewport=0.0cm 0.0cm 19.0cm 8.0cm]{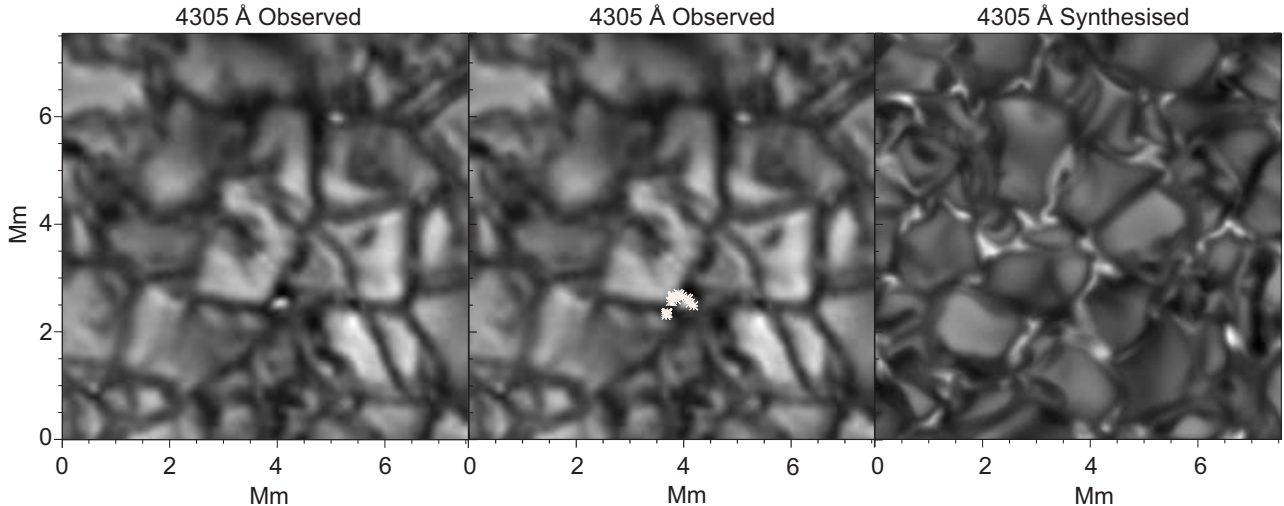} 
\caption{The \textit{left} panel displays the G-band region enclosed in the red box in Fig.~\ref{fig:rosafov}. The 
motion of an MBP flows derived with a point and click method is highlighted by the over-plotted stars 
(\textit{middle panel}). The motion is from left to right.  The \textit{right} image is the 
synthesised G-band from the MURaM code.}\label{fig:photflows}
\end{figure*}

\section{THE MAGNETIC STRUCTURE}
The region of the Sun observed by ROSA is a typical quiet Sun area. The focus of the investigation will be on a 
sub-region of the ROSA field of view that contains a feature of particular interest. Images of this sub-region are 
shown in Fig.~\ref{fig:rosafov}. In the G-band images, a number of MBPs are clearly identifiable. The bright 
points are a good proxy for strong concentrations of magnetic flux, however, strong magnetic flux can still be 
present without an associated MBP (\citealp{BERTIT2001}). Specific conditions appear to be needed for an MBP to 
appear in G-band (e.g., \citealp{CARetal2004}; \citealp{ISHetal2007}). On viewing the movie of the
G-band time series, the MBPs are seen to be pushed and jostled by the convective motions of the granules, merging 
and splitting and appearing and disappearing from the bandpass. 

The Ca K images (Fig.~\ref{fig:rosafov}b) also show bright points, displaying a close mapping to the MBP's seen in 
G-band. This suggests the Ca K bright points are the chromospheric counterpart to the G-band MBPs. However, the 
Ca K bright points are more diffuse,  possibly due to expansion of the magnetic flux with height and partially due to 
the properties of the filter. The bright points are also more persistent than the G-band MBPs, hence show a different 
morphology. The images also contain acoustic grains and reversed granulation, which have been identified 
previously. These features {(possibly including the diffuse MBP nature)} are present due to the large width of 
the filter bandpass used for these observations, 
meaning that the Ca K images here have contributions from both photospheric and chromospheric sources (for a 
detailed explanation, see \citealp{LEEetal2006}). Selecting a sufficiently narrow filter to image Ca K line centre can 
avoid the photospheric contamination and allow mainly chromospheric contributions (see, e.g., 
\citealp{CAUetal2008}; \citealp{REAetal2009}; \citealp{PIEetal2009}). 

In the H$\alpha$ core bandpass (Fig.~\ref{fig:rosafov}c), few signatures of the MBPs are visible. The region is 
covered with fine-scale structure, such as elongated, cell-spanning fibrils and a few shorter mottles (jet-like 
structures). Due to the narrow width of the filter, the H$\alpha$ core images show the {\lq true \rq}, magnetically 
dominated (low-$\beta$) chromosphere (\citealp{RUT2006,RUT2007}). The observed features, i.e., the fibrils and 
mottles, are thought to highlight the chromospheric magnetic field, with the structures likely to be density 
enhancements along individual magnetic field lines (this is supported by recent modelling of  
H$\alpha$ line formation by \citealp{LEEetal2012}).

In the following section we pay particular attention to the area highlighted by the dashed box in 
Fig.~\ref{fig:rosafov}. A number of fibrils and mottles can be seen to emanate from this region. At the photospheric 
level, the G-band images reveal that a number of MBPs appear and disappear. The counterpart Ca K bright points 
remain present, although some of the brighter elements fade and brighten with time. 
These underlying features lead us to suggest the highlighted H$\alpha$ region is the chromospheric section of a 
of a magnetic flux concentration, associated with an MBP, that has undergone significant expansion from the 
photosphere. The observed MBPs diameters are 
$\sim100-200$~km while the H$\alpha$ feature has a diameter of $\sim3000$~km. The described magnetic 
structure is thought to be common in the quiet solar atmosphere (e.g., \citealp{DOWetal1986}; \citealp{PET2001}) 
and is also revealed in observations of chromospheric swirls by \cite{WEDROU2009} and \cite{WEDetal2012}.

The magnetic structure above the chromosphere is much harder to determine. Fig.~\ref{fig:rosafov} (d-f) displays 
images from SDO/AIA in the $304$~{\AA}, $171$~{\AA} and $193$~{\AA} bandpasses. A near-by active region is 
visible in the upper right hand corner of each image, however, it can be seen that the ROSA field of view is far from 
the active region in an area of quiet Sun.


\begin{figure*}[!tp]
\centering
\includegraphics[scale=0.67, clip=true, viewport=0.0cm 1.0cm 20.5cm 28.5cm]{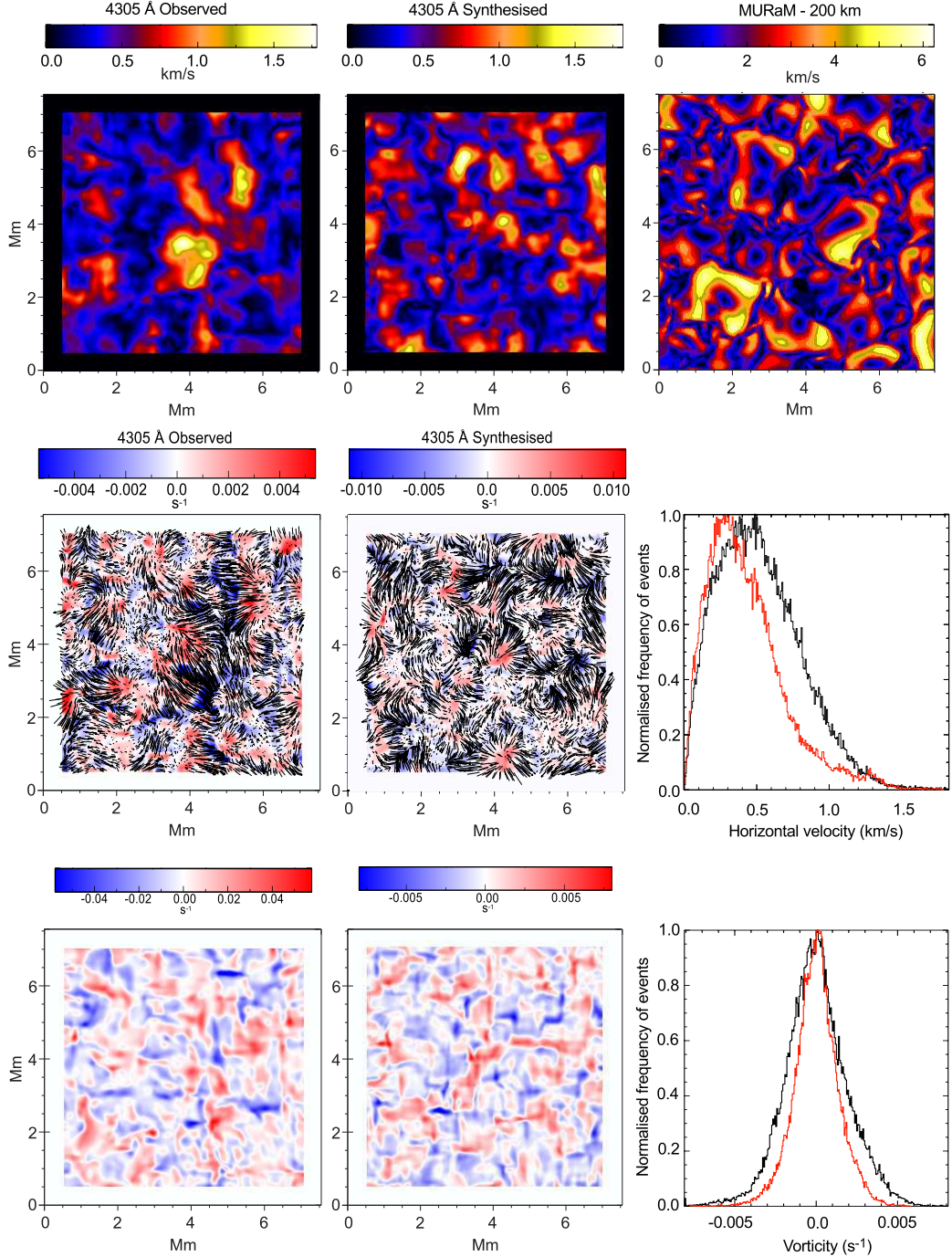} 
\caption{The top row shows the magnitudes of velocities of photospheric flows as determined from the observed G-
band (\textit{left}), the synthesised G-band (\textit{middle}) and the MURaM code at a geometric height of 
$200$~km above the continuum formation level (\textit{right}). The velocities are averaged over 
30 frames ($\sim230$~s). The \textit{middle} row displays velocity vectors and divergence of photospheric flows. 
The arrows display the averaged velocity vectors determined with LCT from the observed (\textit{left}) and simulated 
(\textit{middle}) G-band. A histogram (\textit{right}) showing the distribution of the horizontal velocity for the observed (red) and the 
simulated (black) data. The \textit{bottom} row displays vorticity ($s^{-1}$) calculated from the horizontal 
photospheric flows. The panels displays the averaged vorticity from the observed (\textit{left}) and simulated (\textit{middle}) G-band. A histogram (\textit{right}) showing the distribution of the vorticity for the observed (red) and the simulated (black) data. }\label{fig:pf_mag_vel}
\end{figure*}


\begin{figure*}[!htp]
\centering
\includegraphics[scale=0.9, clip=true, viewport=0.0cm 0.0cm 18.0cm 17.0cm]{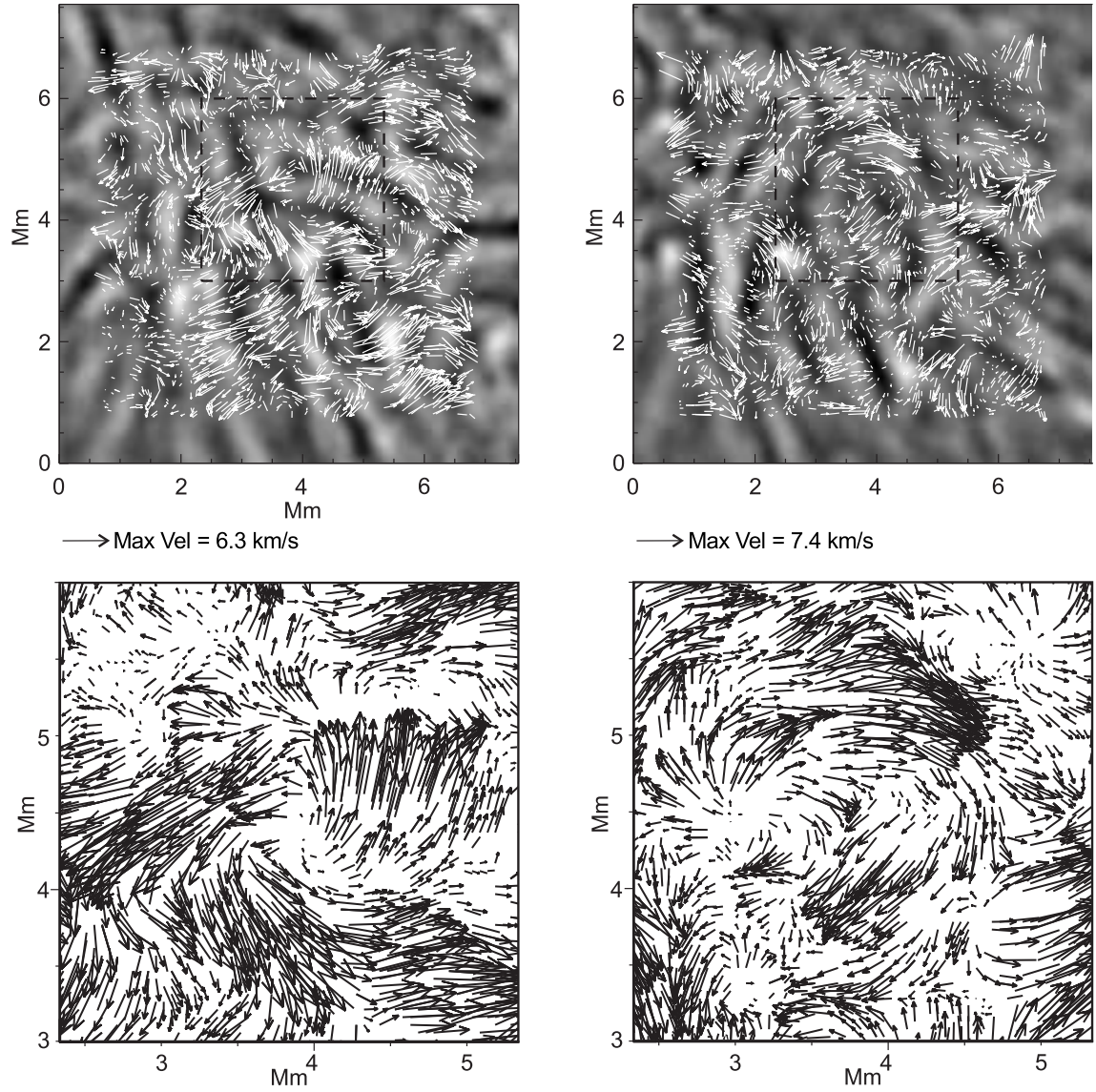} 
\caption{The chromospheric response to photospheric vortices. Torsional motion is observed in H$\alpha$ and 
demonstrated by applying LCT to frames between $t=207-277$~s (left) and $t=610-690$~s (right). The torsional 
motion is centred approximately on the point located at (3.8,4.5)~Mm. The vectors are over plotted on the unsharp 
masked and atrous filtered $H\alpha$ data at $t=200$~s (left) and $t=600$~s (right). The region shown 
corresponds to the dashed box in Fig.~\ref{fig:rosafov}c. The bottom row displays enlarged versions of the regions 
centred on (3.8,4.5)~Mm (as marked by the boxes). The density of vectors has also been increased by a factor of
two and a half. }\label{fig:chromflows}
\end{figure*}

\begin{figure*}[!htp]
\centering
\includegraphics[scale=1.0, clip=true, viewport=0.0cm 0.0cm 12.5cm 9.8cm]{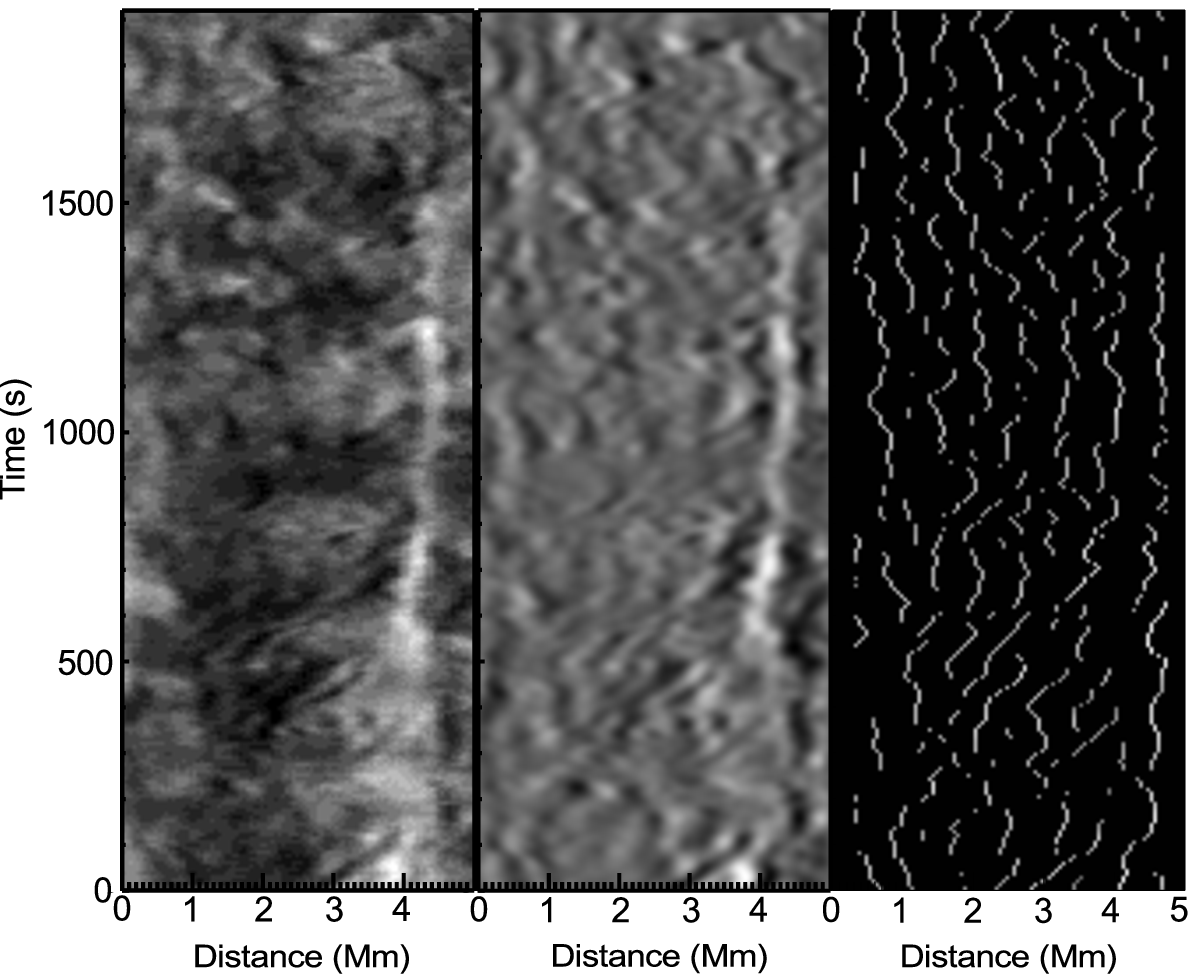} 
\caption{Overview of the technique for identifying transverse waves. The \textit{left} panel shows time-distance plot 
of a cross-cut taken perpendicular to the fibrils (i.e., perpendicular to the dotted blue line in Fig.~\ref{fig:rosafov}c). 
The \textit{middle} panel is the same cross-cut that has been subject to unsharp mask and with the high frequency 
component removed. The \textit{right} panel displays the fibril paths obtained from the time-distance plot in the 
middle panel. The lines highlight the dark features.}\label{fig:skel}
\end{figure*}

\section{FLUX TUBE DYNAMICS}
Now, our attention is focused on the dynamical behaviour of the unique region identified in the previous section 
(dashed box Fig.~\ref{fig:rosafov}). To provide insight into the horizontal motions of the magnetic elements we 
exploit LCT. This technique was originally introduced into solar physics to track large-scale (spatial and temporal) 
photospheric flows, e.g., \cite{NOVSIM1988}, and the technique has been developed further for use on 
magnetograms by incorporating the induction equation (e.g., DAVE - \citealp{SCH2006}). 

We developed an LCT algorithm based on the routine $tr\_get\_disp.pro$ (developed by T. Tarbell) available in the 
SSW TRACE software tree. A recent 
version of this routine first apodizes a set of images using a Hanning function, then computes the cross-correlation 
between the images. The position of the maximum of the cross-correlation corresponds to the integer pixel shifts 
between the images. A local region centred on the maximum is then fitted with a quadratic polynomial regression 
equation to locate the maximum with sub-pixel accuracy, hence providing sub-pixel accuracy on the shifts. 

The routine was tested on sample data, which consists of hundreds of different images taken from the ROSA data 
series. The images were shifted using the routine $shift\_image.pro$ (available in the general SSW software tree). 
Taking averages over large numbers of images, we find that shifts as small as $0.03$ pixels can be resolved to 
within $\pm0.005$ pixels. The accuracy of the LCT improves with increasing shift size. {There appears to 
be potential for great accuracy when calculating small shift values statistically, however, a more rigorous test of 
$tr\_get\_disp's$ ability to resolve shifts is needed.}

From the derived shifts we are able to determine the horizontal photospheric flow, $U_h$. Further, we can 
calculate the divergence of the velocity, $\nabla\cdot U_h$, and the photospheric vorticity, $\nabla\times U_h$. 

\subsection{LCT on G-band data}
The G-band data series has a very high cadence at 0.96~s. However, the previously reported values for 
photospheric flows are $1-2$~km\,s$^{-1}$, hence the expected shift between each frame is $<0.03$ pixels. To 
improve accuracy, every eighth frame is then selected from the G-band series, increasing the 
cadence to 7.68~s and expected shifts to $>0.08$ pixels ($0.5$~km\,s$^{-1}$). Further, a sonic filter is applied to 
the data to suppress the influence of p-modes and stochastic variations in intensity. The data is then re-sampled 
using linear interpolation to achieve a pixel size of $25\times25$~km$^2$.

The LCT is performed on frames chosen from the first sixty frames because the seeing during this period is 
relatively stable. For each data set that we apply LCT to, the LCT is preformed on many sub-regions of the overall 
image, with window sizes determined by the dominant features at that wavelength.  A window size 
of $40\times40$~pixel$^2$ ($1\times1$~Mm$^2$) is used such that features on the order of the 
granulation will contribute to the cross-correlation. This, along with the apodization of the 
window, will reduce the effect of large intensity gradients on resolving shifts.

The $7.5\times7.5$~Mm$^2$ boxed region in Fig.~\ref{fig:rosafov}a is selected (close-up of this region is shown in 
Fig.~\ref{fig:photflows} left panel) and subject to the LCT algorithm. The photospheric flow is calculated over a 30 
frame average, revealing long-lived flow patterns. 

The magnitude of the photospheric velocities are shown in Fig.~\ref{fig:pf_mag_vel}. A histogram plot of the 
velocity distribution (Fig.~\ref{fig:pf_mag_vel}) is in line with results from MBP tracking, e.g., \cite{KEYetal2011};  
\cite{CHITetal2012}, and from long-term flow measurements, e.g., \cite{NOVSIM1988}, \cite{BERetal1998}. The 
velocity vectors and divergence are also shown in Fig.~\ref{fig:pf_mag_vel}. The directions of the velocity vectors 
results are consistent with the flow direction as seen by a visual inspection of the G-band movie. Evidence for 
vortices are present, with the velocity vectors marking out curved paths or arcs thought to be demonstrating the 
vortex behaviour. This is verified by calculating the vorticity from the horizontal velocity, which demonstrates that 
vorticity is present across the the photospheric region (Fig.~\ref{fig:pf_mag_vel} - bottom row, left panel). A 
histogram of vorticity values is given in Fig.~\ref{fig:pf_mag_vel} and the values obtained here are in agreement with 
previous observational measurements (e.g., \citealp{BRAetal1988}; \citealp{VARetal2011}).

In addition, the motion of an individual MBP is tracked over twenty frames using a crude (but efficient) point and 
click method. The MBP selected lies directly under the chromospheric feature, which suggests it is the photospheric 
counterpart.  The tracking of the individual bright point reveals the MBP 
follows an arc-shaped path across the photosphere, demonstrating the MBP is in 
the presence of a photospheric vortex. This is confirmed as the MBP is also seen to lie in a region of strong, 
negative vorticity in Fig.~\ref{fig:pf_mag_vel}.

\subsection{LCT on simulated G-band data}
To validate the results of the LCT technique, we performed tests on a numerically simulated $G$-band image 
sequence. The synthetic data were produced using radiative diagnostics of the three-dimensional numerical model 
of magnetised lower solar atmosphere (i.e. convection zone and photosphere) generated by MURaM radiative MHD 
code (\citealp{VOGetal2005}). The domain size chosen is $12 \times 12 \times 
1.4~\mathrm{Mm}^3$ in the horizontal and vertical directions respectively, and the horizontal resolution is 
$25~\mathrm{km}$. The detailed $G$-band radiative diagnostics was carried out using the radiative transport 
code, designed for computationally efficient wide band filter calculations (see, \citealp{JESetal2012b} for more 
details).  We have generated 800 $G$-band images of solar granulation with the average vertical 
magnetic flux of $200~\mathrm{G}$ {(the value of magnetic field strength is consistent with recent observations, 
e.g., \citealp{OROSUA2012}).} The time cadence for the image sequence is about 2 seconds and not constant 
due to variation of the time step in the numerical simulation.

To provide a direct comparison with the observational analysis described in the previous section, the 
resolution of the simulated data is reduced to $50$~km, using every fourth frame to achieve a similar cadence. We 
apply the sonic filter and re-sample using linear interpolation to $25$~km. The 
LCT results are averaged, again, over 30 frames. The results of LCT on the simulated $G$-band image 
sequence demonstrate qualitatively similar plasma flow patterns in both the granular and intergranular 
regions to the observed data (Fig.~\ref{fig:pf_mag_vel}). The results from the 
simulated data show slightly greater values of divergence and vorticity. The absolute values 
of velocity and vorticity in the simulated data are plotted in a histogram (Fig.~\ref{fig:pf_mag_vel}) for direct 
comparison between the 
observed and synthesised G-band images. The distributions can be seen to be comparable although slight 
differences exist between the observed and simulated values. This can be attributed to the fact that the observed 
and simulated regions are inherently different regions, with an unknown quantity of magnetic flux in the observed 
region. Some of the differences may also lie in the quality of the observed data. A measure of the difference in 
quality is the root mean squared contrast. The synthesised G-band images have a mean RMS of $0.17\pm0.003$, 
while the observed G-band images have a much lower mean RMS of $0.11\pm0.02$ and much greater variation in 
RMS between frames. 

The flow speeds in the simulations have also been measured at the geometrical height of approximately 
$200~\mathrm{km}$ above the continuum formation level. They show the presence of the flow features similar to 
those detected by LCT applied to the simulated $G$-band images (Fig.~\ref{fig:pf_mag_vel}). 
However, due to measurement of the flow speeds at the constant geometrical height instead of at optical depth 
corresponding to the $G$-band radiation formation, only a qualitative comparison of these speeds with the 
speeds obtained by LCT is valid and possible.

The simulations are known to be replete with vortex motions. The agreement seen between LCT results performed 
on the observed and simulated data gives us confidence that {the photosphere} also contains numerous 
vortex structures. Torsional Alfv\'en waves have been shown to be excited at the photospheric level in a number of 
numerical simulations. They are excited in magnetic photospheric vortex structures by random motions, such as 
granular flows (see, for example, numerical simulations of photospheric magneto-convection, e.g.  
\citealp{VANBALLetal2011}; \citealp{KITetal2011}; \citealp{SHEetal2011b}, \citealp{SHEetal2012b,SHEetal2012}; 
\citealp{MOLetal2012}), and by synthetic photospheric torsional (vortex-type) drivers (e.g., 
\citealp{FEDetal2011,FEDetal2011c}, \citealp{VIGetal2012}) in the idealised simulations of solar magnetic flux tubes. 

We note the magnetic vortex structures significantly differ in their nature and dynamics from the non-magnetic 
vortices (see, e.g., \citealp{STENOR1998}, \citealp{KITetal2012}). They, and the torsional Alfv\'en waves they 
produce, generate large amounts of Poynting flux (see, e.g., \citealp{FEDetal2011b}, 
\citealp{SHEetal2012b,SHEetal2012}). As has been shown by  \citet{WEDetal2012}, the magnetic vortex structures 
expand from the photospheric level into the corona. They are suggested channels for electromagnetic 
energy to tunnel into the higher layers of the solar atmosphere 
and, therefore, such structures are considered as potential channels of energy transport for coronal heating.

\subsection{LCT on H$\alpha$ data}
On inspecting the H$\alpha$ movie, a torsional (or rotational) motion is 
observable in the area highlighted with the dashed box in Fig.~\ref{fig:rosafov}c. A chromospheric response to 
photospheric vortices has already been observed in high-resolution observations of the lower chromosphere 
(\citealp{WEDROU2009}, \citealp{WEDetal2012}). So far, the reported chromospheric swirl events 
appear to demonstrate rotation in \textit{one} direction only. Here, the H$\alpha$ region appears to 
first rotate in one direction and then rotate in opposite direction quasi-periodically. Such behaviour has been 
observed in the intergranular vortices in realistic simulations of magneto-convection \cite{ 
SHEetal2012b,SHEetal2011}, as well as in the idealised simulations of  flux tubes in the quiet Sun 
\citep{FEDetal2011c}.  This would suggest the presence of torsional Alfv\'en waves, which have also been identified 
previously in chromospheric features situated above MBPs (\citealp{JESetal2009}).  

To demonstrate this behaviour, LCT is applied to the H$\alpha$ data. Before 
performing LCT, the unsharp-mask (USM) technique is applied to the full field of view and the mask is subtracted. 
The mask is created using a box-car average of width 20 pixels. This technique has two benefits: \textit{(i)} the 
removal of large spatial-scale intensity perturbations which could influence results; \textit{(ii)} an 
increased contrast between chromospheric features and the background. The intensity perturbations 
(\citealp{MORetal2012c}) are removed from the time series because they are coherent over a large spatial-scale. 
Further, an atrous spatial filtering algorithm (e.g., \citealp{STAMUR2002}) is applied to each frame. The highest 
frequency component from the spatial filtering contains mainly noise, which is then subtracted to improve the 
signal to noise in each image. 

The torsional motion appears quasi-periodic with an apparent period between 120-180s, hence, the LCT is 
averaged over sets of 9 consecutive frames. {Note, the period range given is based on the apparent rate at which the 
motion appears to change its direction of rotation. The time-series shows the rotation change directions a number 
of times over it's duration. This is calculated by eye from viewing the H$\alpha$ time-series and we do not 
consider it a rigorous measurement of the period.} The velocity vector plots in 
Fig.~\ref{fig:chromflows} confirm the visual impression of torsional motion. 
The first vector plot is averaged over the frames in the range $207-277$~s and the second 
vector plot is averaged over $610-690$~s. These time ranges are used as they 
correspond to the times when the torsional motion is 
most evident. The torsional motion is centred on the region around $(3.8,4.5)$~Mm in the plots and a zoom of this 
region (Fig.~\ref{fig:chromflows} - bottom row) reveals the signature of the torsional motion. The maximum values 
of the velocity amplitude of the torsional motion are $\sim7$~km\,s$^{-1}$ with an average and standard deviation 
of $1.8\pm1.4$~km\,s$^{-1}$. These values are essentially the time-averaged velocity amplitudes, while the peak amplitudes are likely greater as the peak value is a factor of $\sqrt{2}$ greater than the time-averaged amplitude.

\section{TRANSVERSE WAVES}

The focus of this work is now shifted to the fine-scale structure that originates in the previously identified 
H$\alpha$ region. The fine-scale structure we are interested in are the dark, absorption 
features, which are mainly elongated, inclined fibrils and a few mottles (nearly vertical structures thought to be 
similar to spicules). On inspecting the movie of H$\alpha$, the fine-scale structure can be identified to 
exhibit transverse motion. The fibrils are rooted in the region where the torsional motion is observed and the 
torsional motion shakes the fibrils footpoints side to side, which drives the transverse waves.  In this section we 
study the transverse waves supported by these structures and demonstrate new techniques for obtaining 
information about the waves.

To model these transverse waves, the assumption is made that the waves are propagating along an over-dense, 
magnetic flux tube embedded in an ambient plasma environment. The speed at which these waves propagate is 
known as the kink speed, which is defined as,
\begin{equation}
c_k=\sqrt{\frac{\rho_iv_{Ai}^2+\rho_ev_{Ae}^2}{\rho_i+\rho_e}}.
\end{equation}
where $\rho$ is the density and $v_A=B^2/(\rho\mu_0)$ is the Alfv\'en speed, $B$ is the magnetic field and 
$\mu_0$ is the magnetic permeability. Here, we follow convention labelling internal and ambient plasma 
parameters with subscripts $i$ and $e$, respectively.

\begin{figure*}[!htp]
\centering
\includegraphics[scale=0.9, clip=true, viewport=0.0cm 0.0cm 18.1cm 12.5cm]{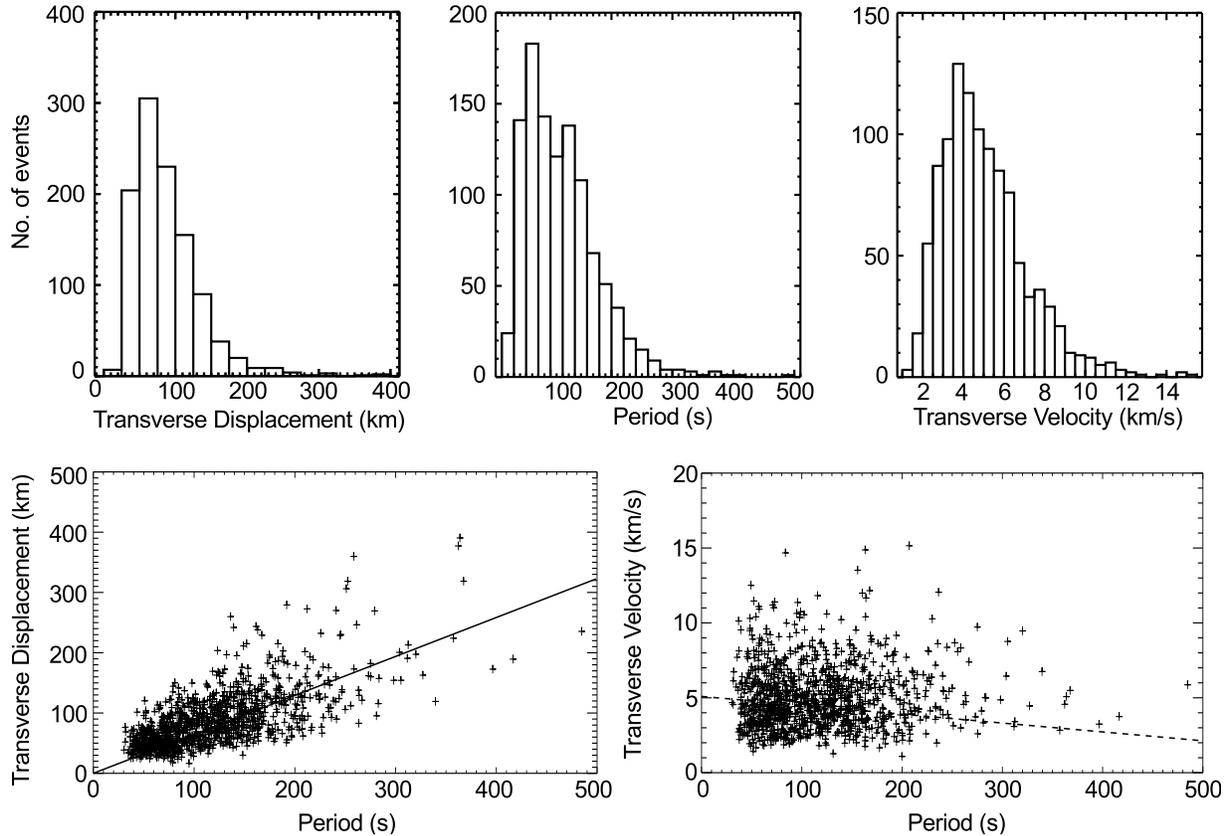} 
\caption{Histograms of measured properties of transverse motions for chromospheric fine structure originating in 
the region displaying torsional motion (\textit{top row}). The 
histrograms show, from left to right, the period, transverse displacement amplitude and transverse velocity 
amplitude. The mean and standard deviations are $P=94\pm61$~s, $A=71\pm37$~km and 
$v_\perp=4.5\pm1.8$~km\,s$^{-1}$. Measured properties of the transverse waves (\textit{bottom row}). 
(\textit{left}) Period versus displacement amplitude. The solid line is a weighted fit to data points that gives a 
velocity amplitude of $\sim4.5$~km\,s$^{-1}$. (\textit{right}) Period versus velocity amplitude. The dashed line is
weighted linear fit to the data points. }\label{fig:hist}
\end{figure*}

\vspace{0.2cm}

The first step in the analysis of the observed transverse waves is to apply the USM procedure to the H$\alpha$ 
images. The orientation of the fine-scale structure is determined (see dotted line over plotted on the H$\alpha$ 
image, Fig.~\ref{fig:rosafov}c) and a series of 42 cross-cuts, separated by 2 pixels, are placed perpendicular to their 
axis. From the 
cross-cuts, time-distance plots are created, an example of which is shown in Fig.~\ref{fig:skel}. The time-distance 
plots clearly reveal the transverse motions of the fibril structures. Next, to reduce the background noise, an atrous 
filtering algorithm is applied to the time-distance plots and the highest frequency component is subtracted. 

The individual, dark fibrils in each cross-cut are then located in the following manner. Cross-sectional flux profiles 
of fibrils have an almost Gaussian shape, with the minimum value of intensity at the centre of the fibril 
cross-section. For each time slice, the pixels with the minimum intensity in a localised region are located. A 
gradient threshold is then applied allowing for isolated fibrils to be located. The gradient of the neighbouring 5 
points on either side of the minimum value pixel is calculated and has to be larger than the threshold value for the 
minimum point to be selected.  Should two fibrils approach each other or 
begin to cross, the gradient becomes shallower around the minima. Optimisation of the threshold level allows for 
a significant number of isolated fibrils to be identified (Fig.~\ref{fig:skel}c), with the fibril paths in time being traced 
out. This method only locates the centre of the structure to within one pixel and hence has an error of 
$\pm50$~km on each point. 

For each fibril path identified at this stage, we then select only the paths that obey certain conditions: (i) the path 
has to be longer than 5 pixels; (ii) The maximum time separation between neighbouring points in the thread has to 
be less than 4 pixels ($30$~s), the {\lq missing\rq} pixels are then recorded as having the same spatial position as 
that of the first of the time-separated neighbours; (iii) Points neighbouring in time cannot be separated spatially by 
more than 4 pixels ($200$~km). This last condition would limit any measured velocity amplitudes to 
$<26$~km\,s$^{-1}$.    

Once we have selected the fibril paths, the results are then fitted with a Levenberg-Marquardt non-linear fitting 
algorithm (mpfit.pro  - \citealp{MAR2009}). A function of the form \begin{equation}
F(t)=G(t)+A\sin(\omega t-\phi)
\end{equation}
is used to fit the oscillations. Here, $G(t)$ is a linear function, $A$ is the displacement amplitude of the oscillation, 
$\omega$ the frequency and $\phi$ the phase of the oscillation. The fitting algorithm is supplied with the errors on 
each data point, where it is assumed the given error is the one-$\sigma$ uncertainty. The fitting routine also 
calculates the one-$\sigma$ error to each fit parameter. The fitted results are kept if the threads are longer than 
$3\cdot2\pi/4\omega$ and the errors on the fitted parameters are smaller than the magnitude of the parameter 
values. Transverse velocity amplitudes for the waves can be obtained from the fit using the relation $v_\perp 
=\omega A$ and errors in the velocity amplitude are calculated by summing in quadrature.

Over the 42 cross-cuts a total of 1100 oscillations are measured. A histogram of the periods, transverse amplitudes 
and velocity amplitudes are given in Fig.~\ref{fig:hist}. A Gaussian fit to the distributions gives means 
and standard deviations of $P=94\pm61$~s, $A=71\pm37$~km and $v_\perp=4.5\pm1.8$~km\,s$^{-1}$. It is 
also interesting to plot period versus displacement amplitude (Fig.~\ref{fig:hist} - bottom row left panel). It is 
apparent that there is a 
direct relationship between transverse displacement amplitude and period. We perform a weighted linear fit to the 
data points, calculated with the Levenberg-Marquardt algorithm with the constraint the line passes through (0,0). 
The result is over-plotted and the derived velocity amplitude from the trend is $\sim4.2$~km\,s$^{-1}$, in 
agreement with the mean value.

{In Fig.~\ref{fig:hist} (bottom row right panel) the velocity amplitude as a function of period is shown. Qualitatively 
the velocity amplitude 
appears relatively constant as a function of period, however, a weighted linear fit to the data points suggest the 
velocity amplitude decreases with period. This implies that the energy, which is {proportional to the square of the 
velocity}, contained within this frequency range of waves (2-30 mHz) is approximately constant as a function of period. It is worth noting that 
the estimates of the temporal power spectra of the horizontal velocity of the photosphere by \cite{MATKIT2010} 
using the G-band filter of \textit{Hinode/SOT} showed a change of trend in the log-log gradients of Fourier power 
vs frequency at 4.7 mHz (approximately 213~s period). For frequencies less than 4.7 mHz they found an $f^{-0.6}$ 
relationship and for frequencies greater than 4.7 mHz it was $f^{-2.4}$. The derived photospheric power spectra, 
found by the method of LCT were taken as a proxy for the actual photospheric driver of transverse waves for a 
numerical simulation by \cite{MATSHI2010} (in their case the $m=0$ torsional Alfv\'{e}n wave and not the $m=1$ 
kink waves we observe here). A number of questions now arise, in comparing our measurements of transverse 
chromospheric velocity amplitude vs period measurements in chromospheric fibrils with the photospheric power 
spectra of \cite{MATKIT2010}. Firstly, our results suggest that \cite{MATSHI2010} may have used the incorrect input 
power spectra, i.e., we see no evidence of wave energy decreasing with decrease in period. This is particularly 
relevant for wave heating since, mechanisms such as resonant absorption, phase mixing and ion-neutral damping 
are all more efficient at higher frequencies. On the other hand, if the photospheric power spectra obtained by 
\cite{MATKIT2010} are a reasonable proxy for the driver of transverse waves, the output is notably different at 
chromospheric heights and must be explained by further study.}

We should note that all fits calculated here are under the assumption that the error in the period can be neglected 
over the error in the transverse displacement and velocity amplitudes. This may be justified as the mean 
error on the period, transverse displacement and velocity amplitudes are $8\%$, $21\%$ and $23\%$, respectively.

\section{DISCUSSION AND CONCLUSIONS}
Tracking wave propagation through the lower layers of the solar atmosphere is an important step in being able to 
assess the viability of the various suggested mechanisms for atmospheric heating and solar wind acceleration. In 
Section.~4 we present observational evidence that photospheric vortices occur in the regions of strong magnetic 
flux concentrations. This is found to be in good agreement with the predictions of advanced models of solar 
magnet-convection. Further, we provide evidence that demonstrates the photospheric vortices can excite torsional 
motions in the chromosphere. This is supported by previous results from various numerical simulations (e.g., 
\citealp{VANBALLetal2011}; \citealp{KITetal2011}; \citealp{SHEetal2011b, SHEetal2012b, SHEetal2012}, 
\citealp{FEDetal2011,FEDetal2011c}, \citealp{MOLetal2012}, \citealp{VIGetal2012}) and previous observations  (e.g., 
\citealp{WEDetal2012}). The observation of the vortex and torsional motions is achieved by exploiting LCT to track 
the motions of both photospheric and chromospheric features and identifying the signatures in the resulting 
velocity vector plots. The chromospheric torsional motion appears periodic here (unlike the uni-directional motion 
identified in swirl events, e.g., \citealp{WEDROU2009}), although we cannot resolve the periodic 
behaviour explicitly. The vector plots demonstrate that the region rotates one way and, at a later time, in the 
opposite direction. A better way to obtain details on torsional waves may be to exploit spectroscopic observations 
(e.g., \citealp{JESetal2009}).

Further, the torsional motion is observed to excite transverse waves in the chromospheric fine-structure whose 
footpoints are rooted in the region. The fine-structure takes the form of elongated absorption 
features that are fibrils and a few shorter features, possibly mottles. The 
transverse velocity amplitudes of the transverse waves are comparable to the velocity amplitudes obtained from 
averaging of the torsional motion. The observations imply that the transverse waves are driven by the torsional 
motion, which in turn is driven by the photospheric vortices. The transfer of energy from photospheric vorticies to 
Alfv\'en waves and chromospheric transverse waves has been demonstrated in numerical simulations of the lower 
solar atmosphere (e.g., \citealp{FEDetal2011b}). 

The measured wave properties also reveal interesting information about the chromospheric waves. In particular, the 
measured velocity amplitudes (Fig.~\ref{fig:hist}) suggest that waves with periods in the range $50-500$~s have 
similar power. However, the results presented here are restricted to incompressible waves confined 
in one particular structure in the chromosphere. A more in depth study taking into account numerous 
chromospheric structures is required before more definite conclusions on this issue can be drawn.
    
Finally, we highlight here that our results have some limitations. The limitations are related to the inability to 
resolve small amplitude ($\lesssim25$~km) transverse waves due to spatial resolution. The demonstrated 
relationship between displacement amplitude and period (Fig.~\ref{fig:skel}) implies that spatial resolution also 
limits the ability to detect short period waves ($\lesssim30$~s).  On the other end of the scale, the known lifetimes 
of the chromospheric structures (i.e., $3-5$~minutes) also limits the observations of longer period waves 
($>200$~s). This is reflected in the low number of observations of wave with longer periods. Some of these 
limitations may be overcome if more advanced techniques are employed for wave tracking, {e.g., calculating 
sub-pixel displacements could reveal smaller amplitude wave phenomena. }

\acknowledgements{The authors would like to thank D. B. Jess, M. Mathioudakis and A. Hillier for a number of 
helpful discussions. Observations were obtained at the National Solar Observatory, operated by the Association of 
Universities for Research in Astronomy, Inc. (AURA) under agreement with the National Science Foundation. We 
thank the technical staff at DST for their help and support during the observations. This work is supported by the 
UK Science and Technology Facilities Council (STFC). RM is grateful to Northumbria University for the award of an 
Anniversary Fellowship and to the Royal Astronomical Society (RAS) for the award of an RAS travel grant. GV grateful 
to the Leverhulme Trust for the award of a fellowship. SS research is supported by the Australian Research Council 
Future Fellowship. RE acknowledges M. K\'eray for patient encouragement and is also grateful to NSF, Hungary 
(OTKA, Ref. No. K83133). }

\end{document}